\documentclass[letterpaper, 10 pt, conference]{ieeeconf}  % Comment this line out if you need a4paper

\IEEEoverridecommandlockouts                              % This command is only needed if 
\usepackage{cite}
\usepackage{amsmath}
\usepackage{algorithmic}
\usepackage{xcolor}
\usepackage{overpic}
\usepackage{tikz}
\usepackage{amsfonts}
\usepackage{algorithm}
\usepackage{subcaption}
\usepackage{graphicx} 
\usepackage{eurosym}

\usepackage{url}
\iftrue   
\usepackage[colorlinks=true]{hyperref}
	\usepackage{xcolor}
	\hypersetup{
		colorlinks,
		linkcolor={blue!90!black},
		citecolor={red!90!black},
		urlcolor={blue!20!black}
	}
\fi
\allowdisplaybreaks

\usepackage{amsthm}

\newtheorem*{theorem*}{Theorem}
\newtheorem{definition}{Definition}
\newtheorem*{definition*}{Definition}

\newtheorem{remark}{Remark}

\newtheorem*{example*}{Example}

\newtheorem*{problem*}{Problem}

\title{\LARGE \bf
    User-centric Vehicle-to-Grid Optimization  with an Input Convex Neural Network-based Battery Degradation Model 
}
\author{Arghya Mallick, Georgios Pantazis, Mohammad Khosravi, Peyman Mohajerin Esfahani, Sergio Grammatico% <-this % stops a space
\thanks{This work is supported by the European Union under the project Drive2X (grant number: 101056934) (\textit{Corresponding author: Arghya Mallick)}.}% <-this % stops a space
\thanks{$^{1}$All authors are with the Delft Center for Systems and Control (DCSC), Department of Mechanical Engineering, Delft University of Technology (TU Delft), Netherlands.   
        {\tt\footnotesize E-mail: \{A.Mallick, G.Pantazis, Mohammad.Khosravi, P.Mohajerinesfahani, S.Grammatico\}@tudelft.nl}}%
}

\begin{document}

\maketitle
\thispagestyle{empty}
\pagestyle{empty}

%%%%%%%%%%%%%%%%%%%%%%%%%%%%%%%%%%%%%%%%%%%%%%%%%%%%%%%%%%%%%%%%%%%%%%%%%%%%%%%%
\begin{abstract}
We propose a data-driven, user-centric vehicle-to-grid (V2G) methodology based on multi-objective optimization to balance battery degradation and V2G revenue according to EV user preference. Given the lack of accurate and generalizable battery degradation models, we leverage input convex neural networks (ICNNs) to develop a data-driven degradation model trained on extensive experimental datasets. This approach enables our model to capture nonconvex dependencies on battery temperature and time while maintaining convexity with respect to the charging rate. Such a partial convexity property ensures that the second stage of our methodology remains computationally efficient. In the second stage, we integrate our data-driven degradation model into a multi-objective optimization framework to generate an optimal smart charging profile for each EV. This profile effectively balances the trade-off between financial benefits from V2G participation and battery degradation, controlled by a hyperparameter reflecting the user prioritization of battery health. Numerical simulations show the high accuracy of the ICNN model in predicting battery degradation for unseen data. Finally, we present a trade-off curve illustrating financial benefits from V2G versus losses from battery health degradation based on user preferences and showcase smart charging strategies under realistic scenarios.

\end{abstract}

%%%%%%%%%%%%%%%%%%%%%%%%%%%%%%%%%%%%%%%%%%%%%%%%%%%%%%%%%%%%%%%%%%%%%%%%%%%%%%%%
\section{INTRODUCTION}

%\IEEEPARstart{T}{he} increasing integration of renewable energy sources in the electricity grid poses several challenges as uncertainty in the autonomous generation of prosumers and  supply from electricity companies due to weather conditions,  often renders supply and demand rules used in the past years outdated.  To circumvent this, 

Electric vehicles (EVs) have shown strong potential to provide flexibility through ancillary services to the electricity grid, acting as virtual energy storage devices \cite{liu2013opportunities,aguilar2024potential}, a concept known as vehicle-to-grid (V2G) \cite{kempton1997electric}.
Multiple studies have been dedicated to proposing smart charging solutions for the batteries of a fleet of EVs connected to a grid retailer or parking lot.  Several works \cite{Deori2018, Deori_conf, Mignoni_2023_ECC, Mignoni2023_CST} formulate smart EV charging/discharging as a game where each EV user acts as a selfish agent, optimizing their cost while satisfying coupling operational constraints. Other works consider stochastic formulations of  EV charging   \cite{Zanvettor2024, Pantazis_2020, Pantazis_2022}.

Despite multiple studies on V2G \cite{brooks2002vehicle,v2g_uk}, regulatory challenges and societal concerns have significantly delayed the adoption of V2G technologies and, as a result, the advantages they can offer \cite{aguilar2024potential}. Of crucial importance for V2G is the voluntary participation of EV users. The study in \cite{van2021factors} revealed that the two main reasons for EV users hesitate to adopt V2G are the uncertainty over financial benefits as well as the potential toll on the vehicle's battery health due to charging and discharging. Given these concerns, there is a pressing need to thoroughly investigate the effects of V2G participation on battery health and the associated financial benefits for EV owners.  Given the significant amount of work dedicated to V2G and smart charging, few studies \cite{bishop2013evaluating,wang2016quantifying,thingvad2021empirical,razi2023predictive,lu2024coordinated} have rigorously investigated the impact of V2G services on EV battery degradation. The results of \cite{bishop2013evaluating} indicate that battery health can deteriorate by participating in V2G services. However, the degradation model proposed does not distinguish between battery cell temperature and ambient temperature. On the other hand, \cite{wang2016quantifying} concludes that the degradation due to V2G participation is negligible when compared to regular driving and calendar aging. However, neither study offers a comprehensive analysis of the trade-offs between the financial benefits of participation in V2G and the associated financial losses due to battery degradation.
In \cite{Scarabaggio_2022}, optimal charging policies are proposed for EVs' batteries in a parking lot with photovoltaics, such that energy requirements are met and battery degradation is minimized according to a given deterministic model.
%Furthermore, these studies do not consider the effects of time-varying charging and discharging power (commonly referred to as smart charging) on battery health during V2G sessions—a critical omission given that a recent study \cite{brinkel2024enhancing} has demonstrated the financial benefits of implementing smart charging for V2G services. 
The authors in \cite{thingvad2021empirical} develop instead an empirical method for measuring battery capacity and demonstrate that one-third of the total capacity fade is attributable to cyclic processes, including V2G usage and daily driving. Finally,  \cite{razi2023predictive,lu2024coordinated} incorporate some user-centric considerations in developing smart-charging algorithms under V2G services,  taking battery degradation into account.   
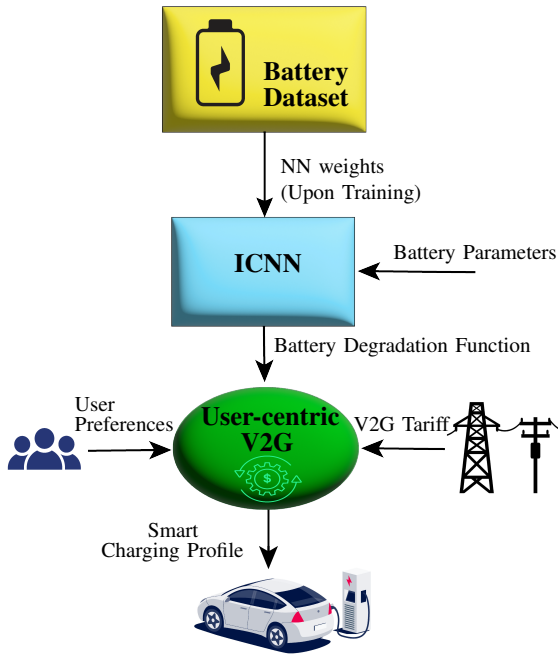
\begin{figure}
    \centering
\begin{overpic}[width=0.42\textwidth]{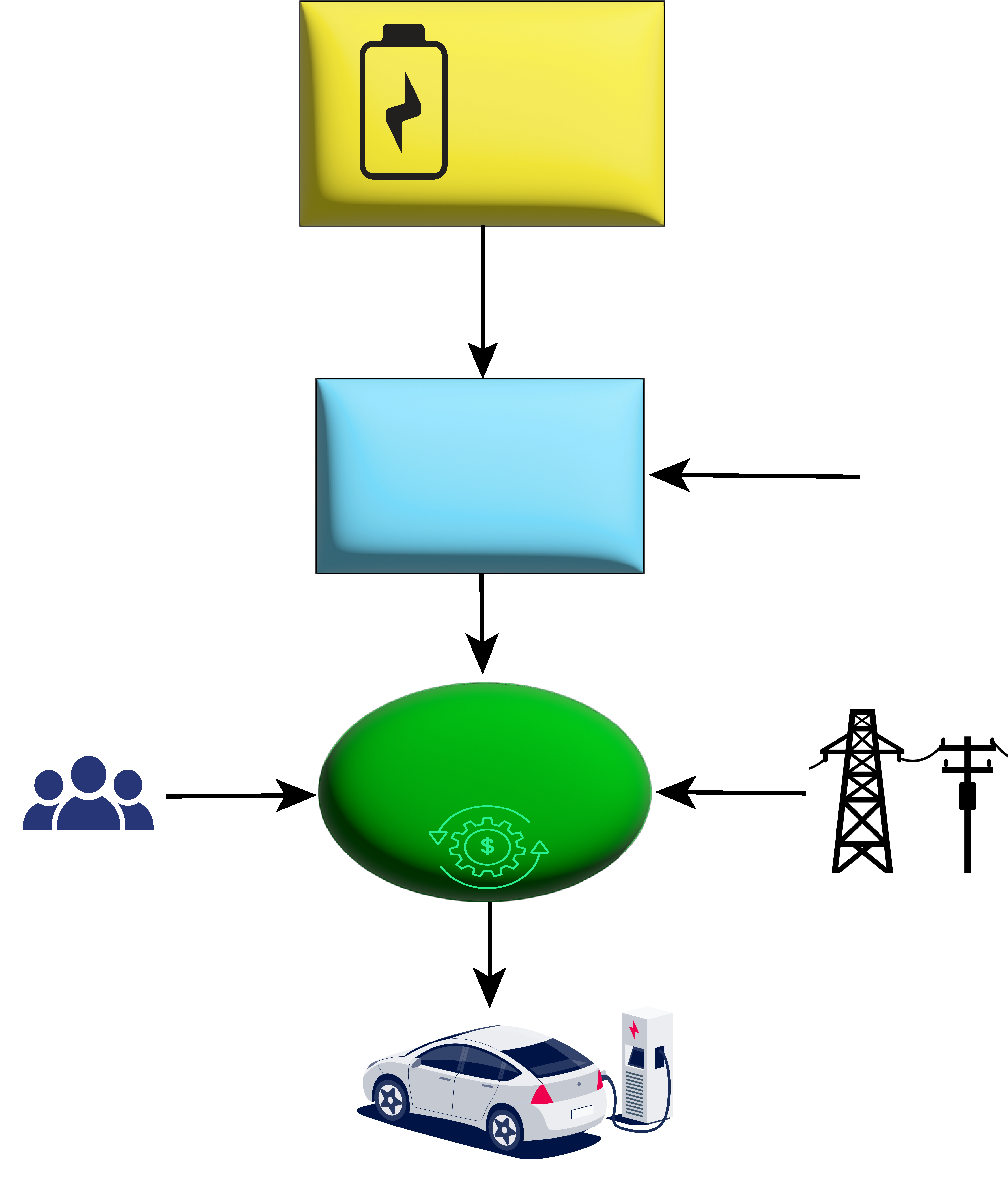}
 \put (31,37) {\textbf{User-centric}}
 \put (37,33) {\textbf{V2G}}
 %\put (50,90) {Battery}
 \put (39,89) { \textbf{Battery}}
  \put (39,85) { \textbf{Dataset}}
  %\put (49,89) {\footnotesize  C-rate}
  %\put (49,86) {\footnotesize Temperature}
  %\put (49,83) {\footnotesize  Time}
  \put (36,60) {\textbf{ICNN}}
  \put (12,39) {\footnotesize User}
  \put (12,36) {\footnotesize Preferences}
  \put (60,62) {\footnotesize Battery Parameters}
  \put (54,36) {\footnotesize V2G Tariff}
  \put (42,48) {\footnotesize Battery Degradation Function}
  \put (23,20) {\footnotesize  Smart}
  \put (16,17) {\footnotesize  Charging Profile}
  \put (43,75) {\footnotesize NN weights} 
  \put (43,71) {\footnotesize (Upon Training)} 
\end{overpic}
    \caption{Pictorial illustration of the proposed data-driven user-centric V2G system.}
    \label{V2G_diag}
\end{figure}

The works above focus mostly on deterministic or empirical battery health degradation models as the basis of numerical experiments using traditional fitting techniques. However, informative datasets have the potential to be used in more advanced machine learning methods to outperform available empirical models for battery degradation and thus enhance the state-of-the-art models with regard to generalization. In recent years, so-called Input Convex Neural Networks (ICCNs) \cite{Amos2017} have gained significant attention in automatic control \cite{ Seel_MPC2022, Bunning2025}  due to their ability to produce a data-driven model, convex with respect to certain inputs\footnote{The term \emph{input} in our setting represents the variables, whose data we use to train the weights of a neural network. If the designer controls an input, it can be viewed as a control variable. If not, it is considered a system parameter.} representing the control variables while retaining the richness of nonconvex relations with respect to other inputs, viewed from a control-theoretic perspective as system parameters. Particularly, in our application, we need the convexity of the battery degradation model with respect to the charging/discharging power of the battery so that the subsequent V2G optimization problem becomes computationally tractable. Therefore, we leverage the richness of ICNNs to model complex battery health degradation phenomena and utilize it for developing a user-friendly V2G optimization framework that enables EV users to \emph{individually} balance their interests between maximizing V2G revenue and minimizing battery degradation. Fig. \ref{V2G_diag} represents an overall schematic of our propositions.  

As such, our main contributions with respect to the related literature are as follows:
\begin{enumerate}
\item \textbf{Data-driven model:} Due to the lack of accurate, generalizable, yet simple, data-driven models for battery degradation, we consider a (partially) input-convex neural network (ICNNs) to model EV battery degradation based on the extensive experimental datasets \cite{Bole2014}. Our ICNN model captures the nonconvex relations between battery degradation and parameters such as battery temperature and time, yet it remains convex with respect to the battery's C-rate. Numerical simulations demonstrate the ICNN's high accuracy in predicting the battery degradation for unseen battery datasets.
\item \textbf{User-centric V2G:} To address the user's risk aversion to their EV's battery health, we use our data-driven convex battery degradation model to formulate a multi-objective optimization problem that captures the trade-off between financial gains from V2G participation and the degradation of battery health. Furthermore, we provide trade-off curves that illustrate the financial losses and benefits based on user preferences and showcase smart charging strategies under realistic price fluctuations.
\end{enumerate}

% The rest of the paper is organized as follows: In Section II we introduce the problem formulation, the notion of Input Convex Neural Network (ICNNs) architectures and describe the proposed  algorithm to solve the problem at hand. Section III shows extensive numerical simulation studies including the problem set-up, the validation of the proposed battery degradation model, the personalized trade-off curves and smart charging profiles for each EV user. Finally, Section IV concludes the paper and presents some research directions which require further research. 
The rest of the paper is organized as follows: In Section II, we review the notation used throughout this paper. Section III introduces the problem formulation as a multi-objective optimization problem. 
To solve the problem, we need to employ the \emph{Input Convex Neural Network} (ICNNs) architectures, which are described in Section IV. The proposed algorithm to solve the problem at hand is introduced in Section V. Section VI provides extensive numerical simulation studies, including the problem set-up, the validation of the proposed battery degradation model, the personalized trade-off curves, and smart charging profiles for each EV user. Finally, Section VII concludes the paper and presents some research directions which require further research. 

%\section{V2G benefits vs Battery degradation}
\section{Notation}
 Throughout the paper, we consider $[ \cdot]_{+}=\max(\cdot, 0)$ and denote the Hadamard product, i.e., the elementwise multiplication between two vectors, by $\odot$.
Let $[T]=\{1, \dots, T\}$ denote the charging horizon of each EV.

%\subsection{Multi-objective Optimization Model}
\section{V2G benefits vs Battery Degradation: A Multi-objective Optimization Approach}
A charging strategy $\text{col}((P_{\text{bat},t})_{t=1}^T)$ should satisfy the following bounds on charging/discharging power of the  EV charger  
    \begin{equation} \label{p_bounds}
        -\overline{P} \leq P_{\text{bat},t} \leq \overline{P}, \: \forall t \in [T],
    \end{equation}
    where $P_{\text{bat,}t}$ is the charging/discharging power of the battery. Furthermore, $P_{\text{bat,}t}$ should follow the battery energy dynamics:
    \begin{equation} \label{e_bounds}
        \underline{E} \leq E_0 + \eta_{\text{avg}}\Delta t\left(\sum_{j=1}^{t} P_{\text{bat},j} \right) \leq \overline{E}, \quad \forall \: t \in [T],
    \end{equation}
    where $E_0$ is the initial energy level of the battery, and $\eta_{\text{avg}}$ is the average charging/discharging efficiency. The final energy level ($E_{\text{des}}$), as set by the EV user, is met with $\epsilon$ tolerance by time $T$ as
    \begin{equation} \label{e_final}
        \left|E_{\text{des}} - \left( E_0 + \eta_{\text{avg}}\Delta t\left(\sum_{j=1}^{T} P_{\text{bat},j} \right)  \right) \right| \leq \epsilon.
    \end{equation}
For compactness, we consider $u=\text{col}((P_{bat}, j)_{j=1}^T)$ and collect all operational constraints in the set $\Omega$ given by:
\begin{equation}\label{feasibility}
    \Omega:= \{u \in \mathbb{R}^{T} \mid  \eqref{p_bounds}, \ \eqref{e_bounds} \ \text{and} \  \eqref{e_final} \  \text{hold}, \ %\text{for all}\, 
    \forall\,t \in [T]\}. 
\end{equation}
We now define the objective function associated with the V2G exploitation as follows:
\begin{align} \label{v2g_cost}
\theta^{(1)}_T (u) := \sum\limits_{i\in [T]} \alpha_i(P_{\text{bat},i}\Delta t) 
\end{align}
with  $\alpha_t$ (in \euro/kWhr) denoting the V2G price. This dynamic price inherently encodes different ancillary services offered to the grid, which include demand response, peak shaving, and congestion management. In this paper, the price is provided by the V2G operators (such as EV charging station operators, parking-lot managers, etc.) based on their local-level power dispatch problem with distribution system operators. The cost function in \eqref{v2g_cost} refers to the net charging cost during the V2G session. $\Delta t$ denotes a fixed time interval.
The objective function associated with battery degradation is given by:
\begin{align}\label{obj_2} \ &\theta^{(2)}_T (u,t_{\text{temp}}) := \zeta \left(\sum\limits_{i\in [T]}  \: Q_{\text{loss},i}(P_{\text{bat},i},t_{\text{temp}})\right)\!, 
\end{align}
where $\zeta=(\gamma \times n_{\text{series}} \times n_{\text{parallel}})$ with $\gamma$ being the effective capacity cost of the battery, $n_{\text{series}}$ and $n_{\text{parallel}}$ are the number of series-connected and parallel-connected cells in the EV battery pack. $t_{\text{temp}} \in \mathbb{R}^{T}$ is the temperature of the battery cell over $T$ intervals, and $Q_{\text{loss},i}$ is the capacity loss of a single battery cell at the $i^{\text{th}}$ interval.

We now model a trade-off between the two conflicting objectives of V2G profit maximization and battery degradation minimization using a multi-objective optimization approach. In this case, a user-defined weight is assigned to each objective, reflecting their relative importance to the user. We frame our problem in terms of a multi-objective optimization problem as follows 
% \begin{align}  
%     \label{mo}
%     \min_{u \in \Omega} \:\:  J(u,T, t_{\text{temp}}):= \: \left[ \rho \theta^1 (u) + (1-\rho) \theta^2 (u, T, t_{\text{temp}})\right], 
% \end{align}
\begin{subequations}\label{J_opt}
\begin{align}  
    \label{mo}
    \min_{u \in \Omega} \:\: J(u,T, t_{\text{temp}}),  
\end{align}
where
\begin{equation}
    J(u,T, t_{\text{temp}}):= \:  \rho \theta^{(1)}_T (u) + (1-\rho) \theta^{(2)}_T (u,t_{\text{temp}}),
\end{equation}
\end{subequations}
and   $\rho \in [0,1]$ is the weight chosen by the user to indicate which objective among V2G revenue and battery degradation health is more favorable.
\section{Input convex neural network-based battery degradation modeling}
The capacity loss model \eqref{obj_2} of the EV battery is, in general, difficult to model empirically because of intricate dependences on the temperature ($t_{\text{temp}}$) and time ($T$), and the fact that different battery chemistries manifest different mathematical models. In this paper, we design a data-driven model on the battery degradation cost $\theta^{(2)}_T(\cdot)$ based on ICNNs. In the following sub-section, we present the key properties of such architectures.

 %$\beta$ (in \euro / kWhr) is the user's discomfort factor which is defined as the cost due to deviation from user's desired level of energy

%In what follows, we first introduce the Neural Network architectures related with this work. 

\subsection{Fully Input Convex Neural Networks}
We first present the architecture of a fully input convex neural network (FICNN) as introduced in \cite{Amos2017}. Let $y \in \mathbb{R}^{n_y}$ be the set of inputs of the neural network, where the function resulting from the output of the neural network is $f(y; \bar{w})$, where $\bar{w}$ represents the parameters of the FICNN. 

\begin{definition}  \label{def:FICNN}
Let $k$ be the number of layers of the neural network, $g_i$ the activation function and the hidden state $z_i \in \mathbb{R}^{n_z}$ for each layer $i$, respectively, 
\begin{equation}
\bar{w} = \left( W^{(y)}_{0:k-1}, W^{(z)}_{1:k-1}, b_{0:k-1} \right), 
\end{equation}
the collection of weights and bias, and $y$ the input. Consider the neural network described by Algorithm \ref{Alg:FICNN}. If each element of $W^{(z)}_{1:k-1}$ is non-negative and for each layer $i=0, \dots, k-1$, $g_i$  are convex and nondecreasing, then the Neural Network is called Fully Input Convex Neural Network (FICNN). 
\begin{algorithm}[t]
\caption{Fully Input Convex Neural Network \cite{Amos2017}} \label{Alg:FICNN}
\begin{algorithmic}[1]
\STATE \textbf{Input}: $\bar{w}$ 
\STATE \textbf{for} $i=0,\dots, k-1$
\STATE $z_{i+1} = g_i \left( W^{(z)}_i z_i + W^{(y)}_i y + b_i \right)$
\STATE \textbf{endfor}
 \STATE \textbf{Output}: $f(y; \bar{w}) = z_k$.
\end{algorithmic}
\end{algorithm}
\end{definition}

By leveraging fundamental results on properties of convex functions \cite{boyd2004}, it can be shown that the output function  $f(y; \bar{w})$ of the FICNN in Definition \ref{def:FICNN} is convex in  $y$. As such, a data-driven model can be derived for the battery degradation cost $\theta^{(2)}(\cdot)$ that is convex with respect to all its inputs. 

A key challenge of FICNNs is the joint convexity they enforce across all input parameters. This constraint significantly limits the hypothesis space, preventing the model from capturing nonconvex behavior in certain parameters while maintaining convexity in others. In our case, this would impose an unnecessary restriction by assuming that battery degradation cost is convex with respect to temperature and time, which does not accurately reflect real-world dynamics. To address this limitation, an extension of FICNNs known as partially input convex neural networks (PICNNs) has been proposed in \cite{Amos2017}. PICNNs allow for selective convexity, enabling our model to preserve convexity with respect to specific parameters—such as the battery C-rate while capturing the inherent nonconvexities associated with temperature and time.
\subsection{Partially Input Convex Architectures}
  Neural Networks with a partially input convex architecture can provide a substantial improvement in representation compared to FICNNs due to the enlarged hypothesis space they allow without sacrificing the convexity with respect to the parameters of interest. 
\begin{definition}\label{def:PICNN} 
Consider $k$-layers, the inputs 
$x \in \mathbb{R}^{n_x}$, $y \in \mathbb{R}^{n_y}$, the weight-bias vector 
\begin{equation}
w =
\left(\tilde{W}_i, \tilde{b}_i, W_i^{(zu)}, b_i^{(z)}, W_i^{(yu)}, b_i^{(y)}, W_i^{(u)}, b_i \right),    
\end{equation}
the hidden layer vectors $\tilde{u}=[u_i, u_i^{(z)}, u_i^{(y)}, u_i^{(u)}]$, $f$ is convex in $y$ but not in $x$, and $z=\text{col}((z_i)_{i=1}^{k-1})$. Consider the iterative procedure described in Algorithm \ref{Alg:PICNN}.
If $W^{(z)}$ is elementwise non-negative, then the procedure
is called a Partial Input Convex Neural Network (PICNN). 
\end{definition}
\begin{algorithm}[t]
\caption{Partially Input Convex Neural Network \cite{Amos2017}} \label{Alg:PICNN}
\begin{algorithmic}[1]
\STATE \textbf{Input: $w$ }
\STATE \textbf{for} $i=1, \dots, k-1$
    \STATE $u_{i+1}=\tilde{g}_i(\tilde{W}_i u_i+\tilde{b}_i)$
    \STATE $u_i^{(z)}=\max(W_i^{(zu)} u_i + b_i^{(z)}, 0)$
    \STATE $u_i^{(y)}=W_i^{(yu)}u_i+b_i^{(y)}$
    \STATE $u_i^{(u)}=W_i^{(u)}u_i+b_i$
    \STATE $z_{i+1}=g_i(W_i^{(z)}(z_i \odot u_i^{(z)})+W_i^{(y)}(y \odot u_i^{(y)})+u_i^{(u)}))$ 
    \STATE \textbf{endfor}
    \STATE \textbf{Output} $f(x, y, w) = z_k$,  $u_0 = x$.
\end{algorithmic}  
\end{algorithm}

The data-driven algorithm in Definition \ref{def:PICNN} can be compactly written as a procedure $ \mathcal{F} (x, y, w)$, where $w$ has been trained based on available data using, e.g., empirical risk minimization or mini-batch approaches.

\begin{remark}
An advantage over the FICNN architecture is that only the $W^{(z)}$ terms are required to be non-negative, and the activation function $g_i(\cdot)$ has to be non-decreasing to retain the convexity of the output $f$ in $y$.
\end{remark}

\section{User-centric Data-driven Smart Charging}
We now propose the Algorithm \ref{Algorithm} for data-driven user-centric smart-charging. 
In Algorithm \ref{Algorithm}, we distinguish between two different stages:
\begin{enumerate}
\item \emph{Offline training of the PICNN}: We first pre-process the input-output data from the dataset \cite{Bole2014} and minimize the training loss of PICNN to obtain parameters ($w^*$) of the trained network. In our problem, $y=u$ and $x=\{T,t_{\text{temp}}\}$ of $f(x,y,w)$.  
\item \emph{User-centric Multi-objective Optimization}: We then fix $w^*$ obtained in step (1) and minimize the cost $J(\cdot)$ over the power rate $u$ using automatic differentiation and projected gradient descent, based on the user's preference $\rho$. Note that $ \tilde{T}$ is set by the user, and $\tilde{t}_{\text{temp}}$ could be approximately taken as the forecast of ambient temperature in $ \tilde{T}$ intervals. 
\end{enumerate}
Finally, we retrieve the smart charging profile $u^*$, which will meet all our desired constraints with a guarantee. Thanks to the convexity, our proposed algorithm converges with a global optimal solution.  
\begin{algorithm}[h]   % Start of the algorithm environment
\caption{Data-driven User-centric Smart Charging} \label{Algorithm}  % Caption for the algorithm
\label{alg:PICNN}  % Label for referencing the algorithm
\begin{algorithmic}[1]  % Start of the algorithmic environment with line numbering
\STATE \textbf{Inputs}: Dataset \cite{Bole2014}, user preference $\rho$
    \STATE Train the PICCN on the dataset and obtain $w^*$.
    \STATE Set $\theta^{(2)}_T(u,t_{\text{temp}})= f(u, T, t_{\text{temp}}, w^*)$
    \STATE Given $(\tilde{T}, \tilde{t}_{\text{temp}})$, differentiate $J(u, \tilde{T}, \tilde{t}_{\text{temp}})$ in $u$.
   % \STATE Given $\rho$ from user solve: 
      \STATE    \textbf{for} $\ell=1,2, \dots$ 
      \STATE     $u_{\ell+1} = \text{proj}_{\Omega}[u_\ell-\tau \nabla_u J(u_\ell, \tilde{T}, \tilde{t}_{\text{temp}}) ]$ 
  \STATE    \textbf{until convergence}
  \STATE \textbf{Output}: $u^\ast=u_k$
\end{algorithmic}
\end{algorithm}

\section{Numerical experiments}
\subsection{Simulation Setup} \label{Sec:Simulation}
 We adopt the V2G tariff profile $\alpha_t$ in \eqref{v2g_cost} from \cite{price_data}. We assume EV users engage in the V2G program $12$ hours in a parking lot. The time interval $\Delta t$ is set to 15 minutes. The energy availability within the battery is constrained by upper ($\overline{E}$) and lower ($\underline{E}$) bounds of 0.9 and 0.2 (in per unit), respectively. We define the user tolerance $\epsilon$ as 0.02 (in per unit), and the desired battery energy level $E_{\text{des}}$ is set to 0.7 (in per unit). The battery pack under consideration has a capacity of 50 kWh, with configuration parameters of $n_{\text{series}} = 83$ and $n_{\text{parallel}} = 94$. The terminal voltage of the battery pack ($V_{\text{bat}}$) is 350 V.

Given the availability of various standardized charging technologies, we specifically focus on bi-directional EV chargers with variable charging/discharging rates. As highlighted in \cite{brinkel2024enhancing}, such chargers offer financial advantages over fixed C-rate charging strategies. Accordingly, we select a Level-2 three-phase on-board charger with a 22 kW power rating ($\overline{P}$), which is currently manufactured by Eaton and other major industry players \cite{yuan2021review}. Market analysis in \cite{mohammadi2023comprehensive} indicates that the battery capacity of most passenger EVs falls within the 50–100 kWh range. We conduct our study based on a 50 kWh battery capacity to ensure broad applicability. For our simulation studies, we assume that a new battery costs $207$ \euro/kWh, based on the Chevrolet Bolt battery pack cost reported in \cite{wentker2019bottom}. When the battery capacity degrades by 30$\%$, it can still be repurposed for second-life storage applications, with a resale value of $45$ \euro/kWh \cite{montes2022procedure}. Consequently, the effective cost of the EV battery is calculated as 
% $\gamma = \frac{(207 - 0.7 \times 45)}{0.3} = 585$ \euro/kWh, 
% $$\gamma = \frac{(207 - 0.7 \times 45)}{0.3} = 585\ \text{\euro/kWh},$$  
\begin{equation}
    \gamma = \frac{(207 - 0.7 \times 45)}{0.3} = 585\ \text{\euro/kWh},
\end{equation}
which corresponds to the parameter used in \eqref{obj_2}.
\begin{figure}
    \centering
    \includegraphics[width=1\linewidth]{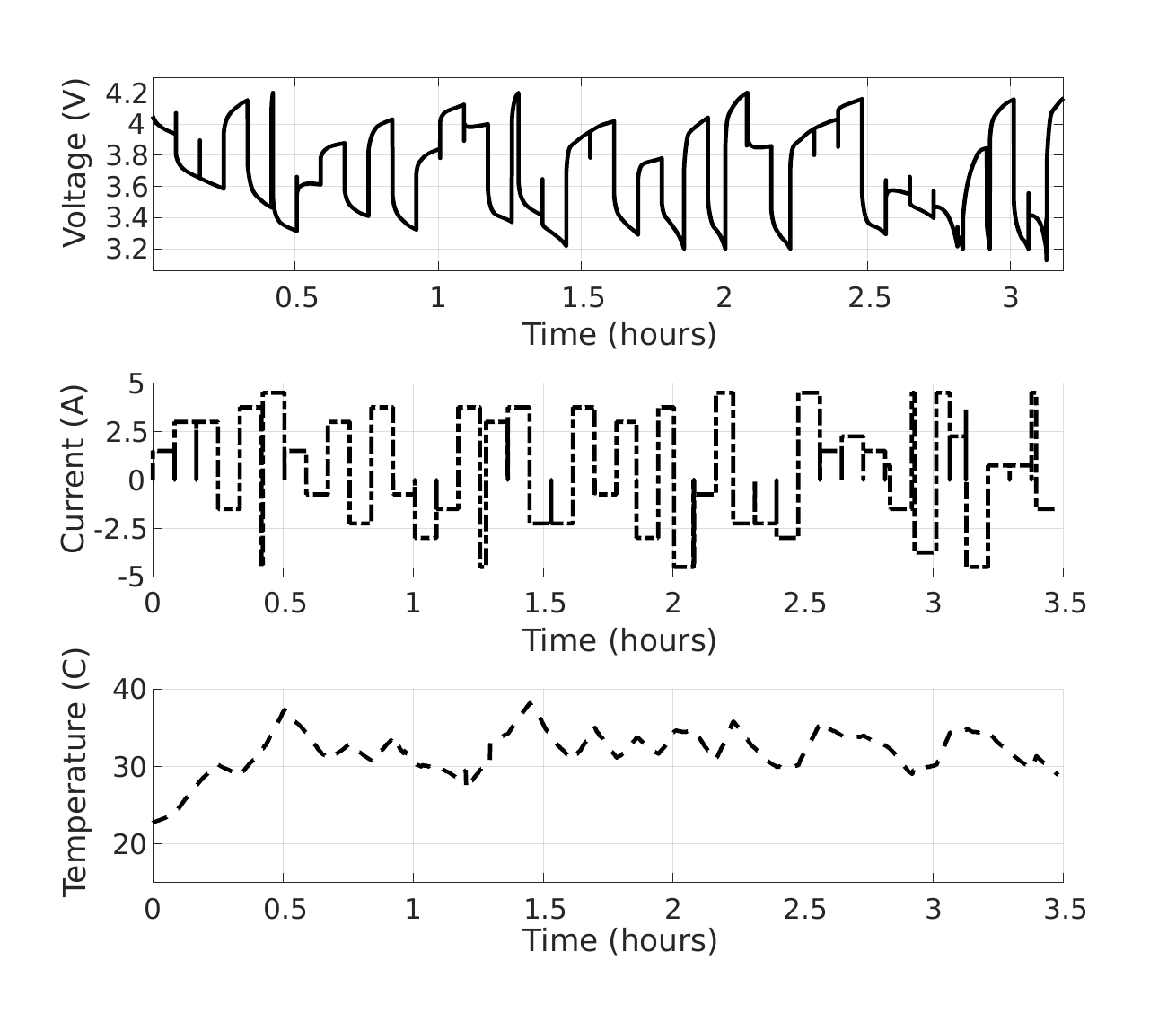}
    \caption{Voltage, current, and temperature plots of RW9 battery cell during randomized battery cycling. The data for the above plot is taken from \cite{Bole2014}.}
    \label{exp_plot}
\end{figure}
\subsection{Battery health prediction results}
\subsubsection{Preparing the experimental dataset}
We use extensive datasets from NASA Ames Prognostics Center of Excellence Randomized Battery Usage \cite{Bole2014} for battery life-cycle tests. The charging-discharging power for each cell is highly randomized in this dataset, which renders this dataset ideal for mimicking practical EV usage. We consider experimental data of four $18650$ Li-ion batteries (identified as RW9, RW10, RW11, and RW12 in the dataset). Although one could define various input features, we primarily focus on key features such as time, battery temperature, and battery cycling C-rate. Fig. \ref{exp_plot} displays voltage, current, and temperature of randomized cycling of RW9 battery cell. Given these inputs, our objective is to predict the battery health degradation in terms of battery capacity fading. Therefore, We train the PICNN using the RW9, RW11, and RW12 battery cell datasets and validate the prediction of capacity fading with an unseen dataset of RW10 battery cell.  
\subsubsection{Battery degradation prediction of PICNNs}
\begin{figure}
    \centering
    \includegraphics[width=0.9\linewidth]{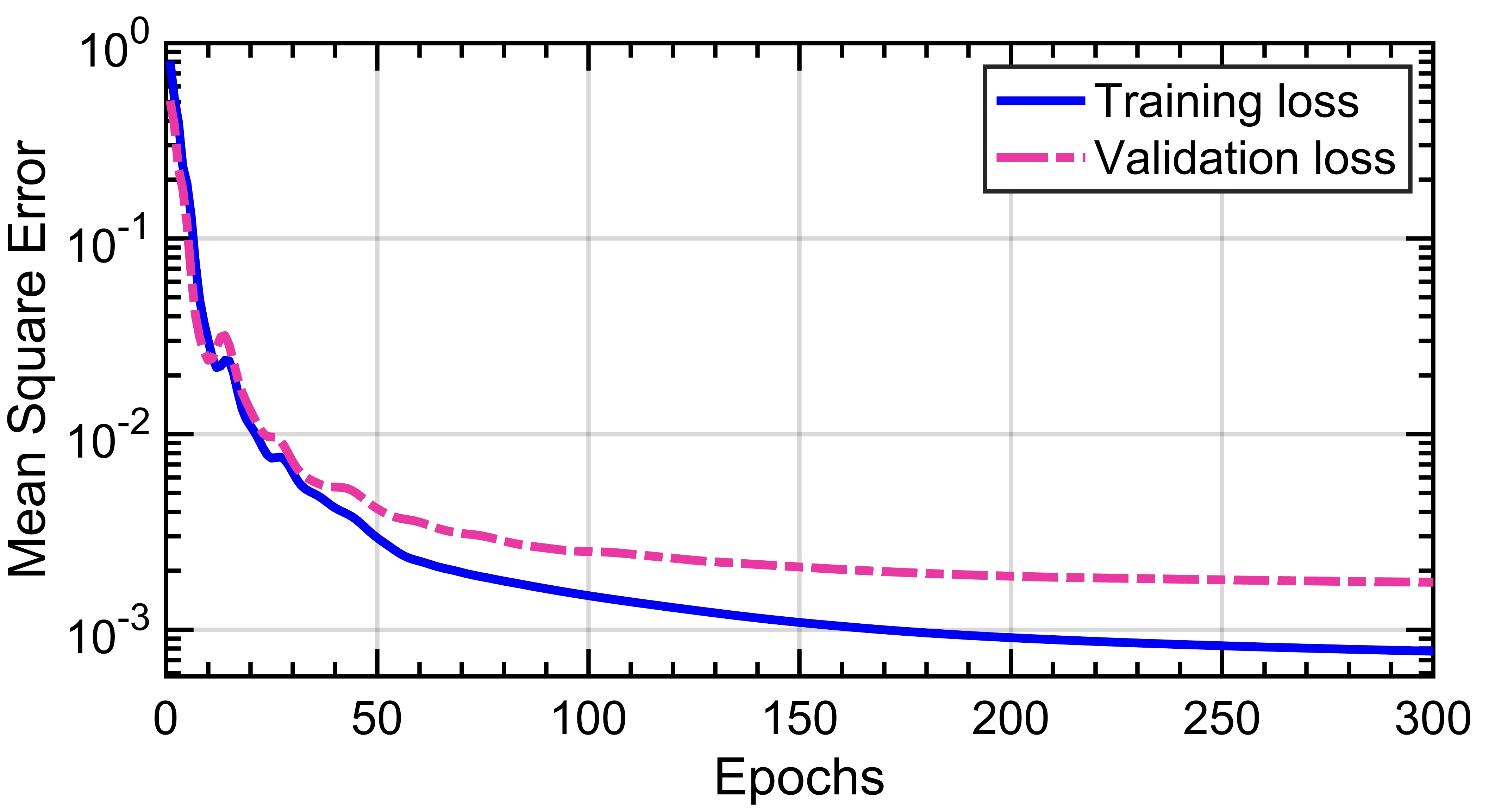}
    \caption{\small PICNN training loss with validation loss.}
    \label{loss_fig}
\end{figure}

\begin{figure}
    \centering
    \includegraphics[width=0.9\linewidth]{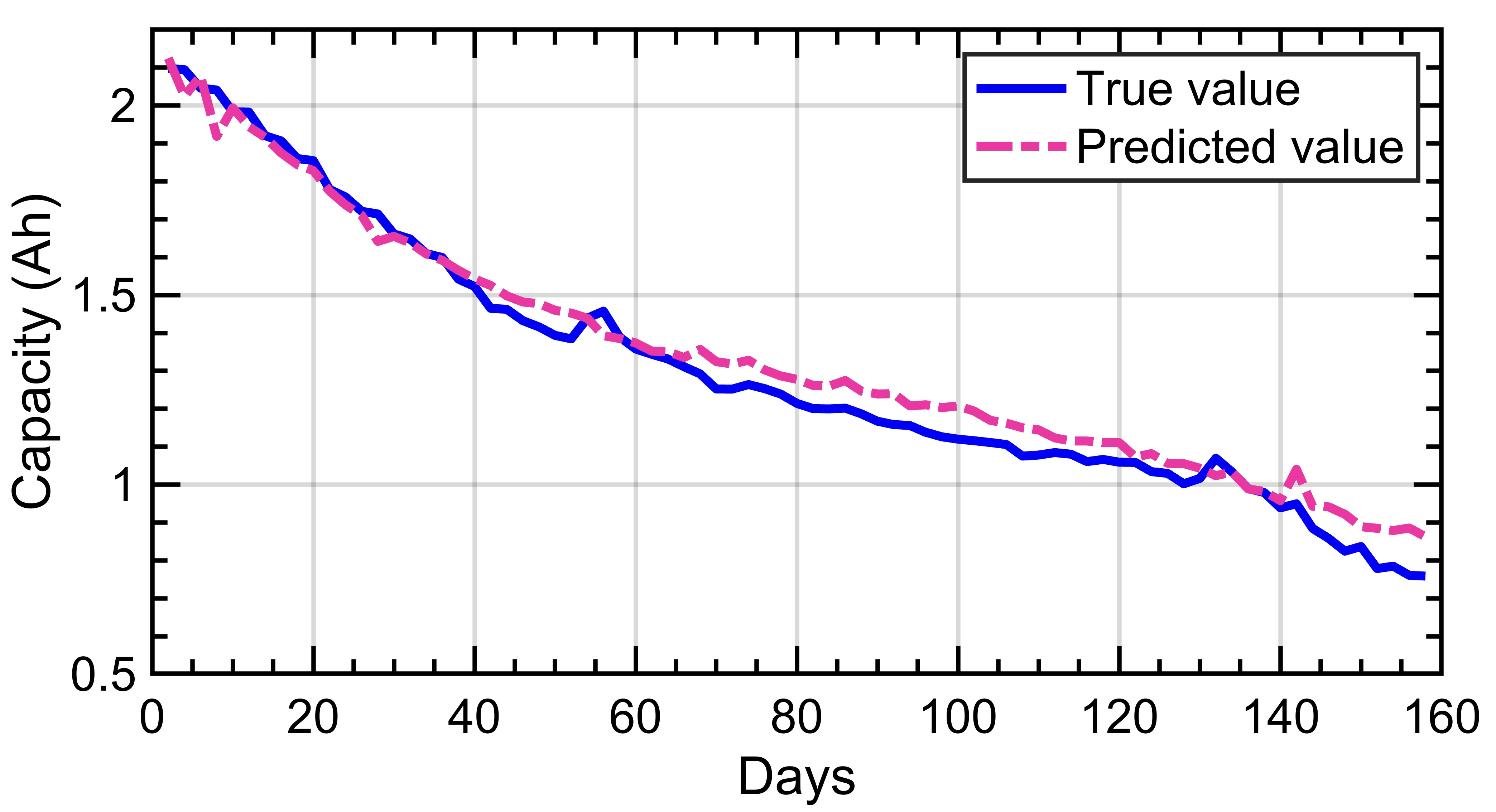}
    \caption{\small Battery capacity prediction vs true capacity comparison.}
    \label{capacity_fig}
\end{figure}

We construct the PICNN with three hidden layers having $32,8,$, and $1$ neurons, respectively. We use the $\mathrm{softplus}$ activation function for the convex part and the $\mathrm{tanh}$ activation function for the non-convex part. The loss function is based on the mean square error (MSE) metric. We optimize our PICNN using an \textsc{adam} solver with an initial learning rate of $0.04$ and decay of $0.1$.  Fig. \ref{loss_fig} depicts the simultaneous training and validation loss of the PICNN. The decreasing trend in the mean square error of both training and validation loss indicates the efficacy of the neural network. In Fig. \ref{capacity_fig}, we show the lifetime capacity prediction of the battery cell (RW10) as an output of PICNN against the ground truth. To measure the goodness of fit of our prediction model, we consider the statistical measure \textit{coefficient of determination} ($R^2$) as the evaluation metric. Consider that the prediction is encoded in $f\in\mathbb{R}^n$ and the ground truth is given by $f^* \in \mathbb{R}^n$. $R^2$ is defined as follows. 
\begin{equation}
    R^2 = \: 1- \frac{\sum_{i=1}^{n} (f^*_i - f_i)^2}{\sum_{i=1}^n (f^*_i-\overline{f}^*)^2} \:,
\end{equation}
where $\overline{f}^*$ is the mean value of $f^*$. An $R^2$ of $1$ indicates that the prediction perfectly fits the ground truth. We find that the value of $R^2$ in our prediction is $\mathbf{0.974}$, which indicates the high performance of PICNNs in predicting the battery capacity fading phenomenon.   
\subsection{Trade-off Curve and Smart Charging Profiles}
\subsubsection{Personalized Trade-off} 
\begin{figure}
    \centering
    \includegraphics[width=0.9\linewidth]{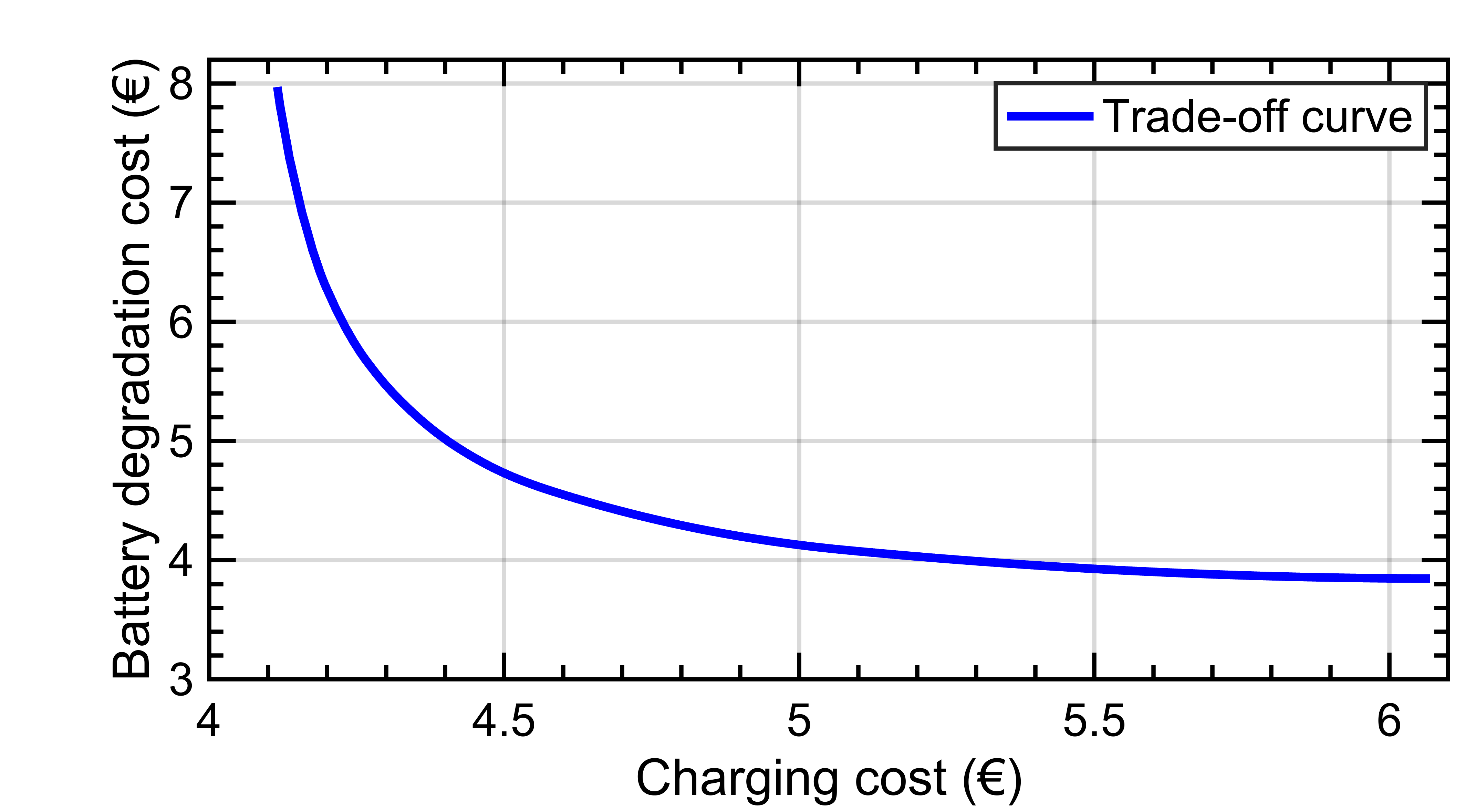}
    \caption{\small Trade-off between charging and battery health degradation cost.}
    \label{trade_off_fig}
\end{figure}

Feeding the degradation cost produced by the ICNN to the multi-objective optimization problem in  (\ref{mo}), we obtain a personalized optimal charging schedule according to the proposed Algorithm \ref{Algorithm}. In Figure \ref{trade_off_fig}, we illustrate the trade-off between the benefits from V2G and battery degradation for different values of the user-defined hyperparameter $\rho$ denoting the subjective importance each user assigns to their battery degradation. The rightmost part of the plot corresponds to the value $\rho=0$, where the user decides to be overly cautious with respect to their EV's battery health. In contrast, the leftmost part of the plot corresponds to $\rho=1$, where the user fully exploits V2G services for financial gain without considering the battery degradation cost. For any other values $\rho \in (0,1)$, the curve provides an informative chart for the EV user with respect to the financial benefits/losses that they will encounter, given their personal preferences and the problem specifications considered in Section \ref{Sec:Simulation}.
\begin{figure}
	\centering
	\begin{subfigure}{\linewidth}
    		\centering
    		\includegraphics[width=0.9\linewidth]{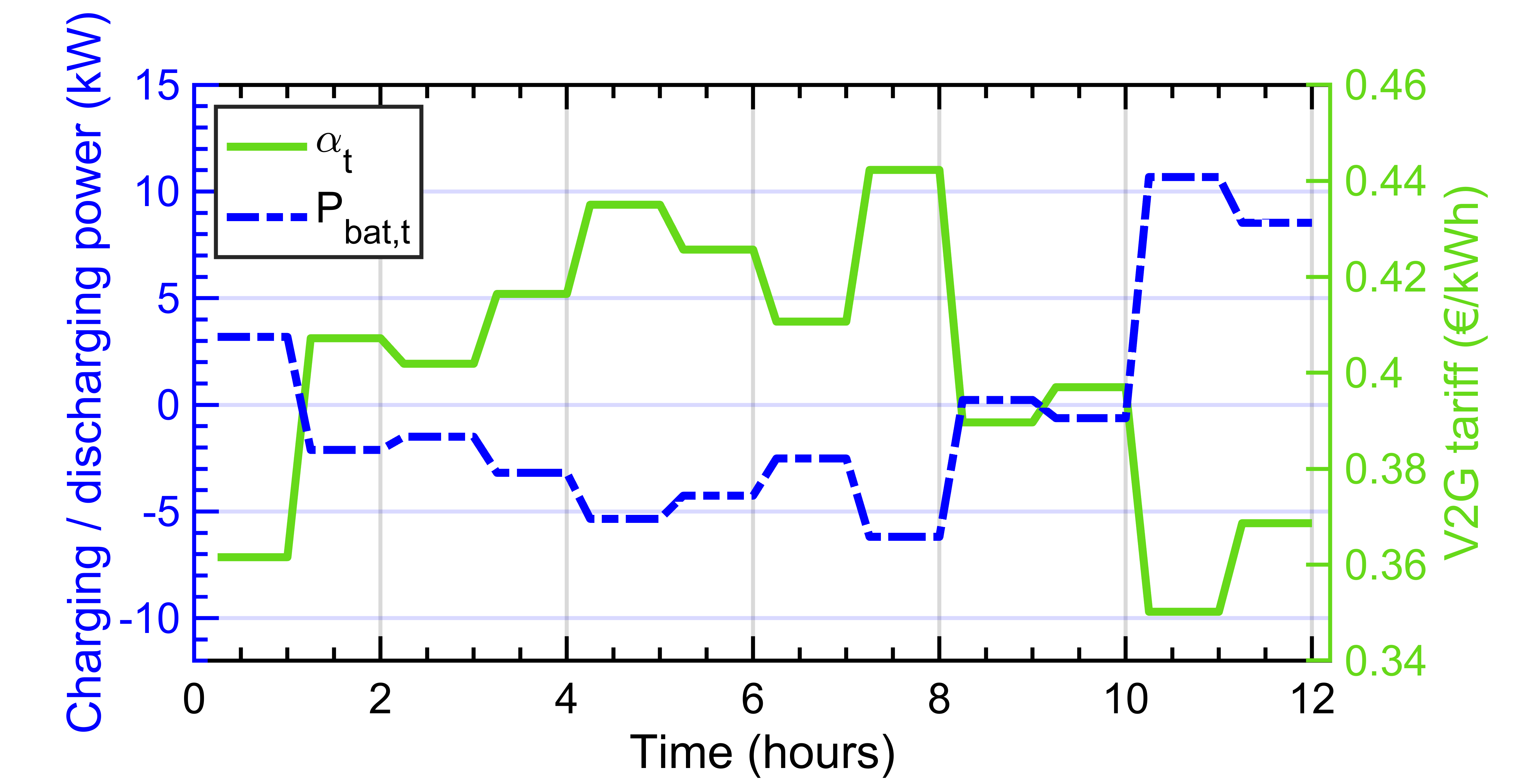}
                \caption{}
    		\label{sensi}
	\end{subfigure}
 
 	\begin{subfigure}{\linewidth}
    		\centering
    		\includegraphics[width=0.9\linewidth]{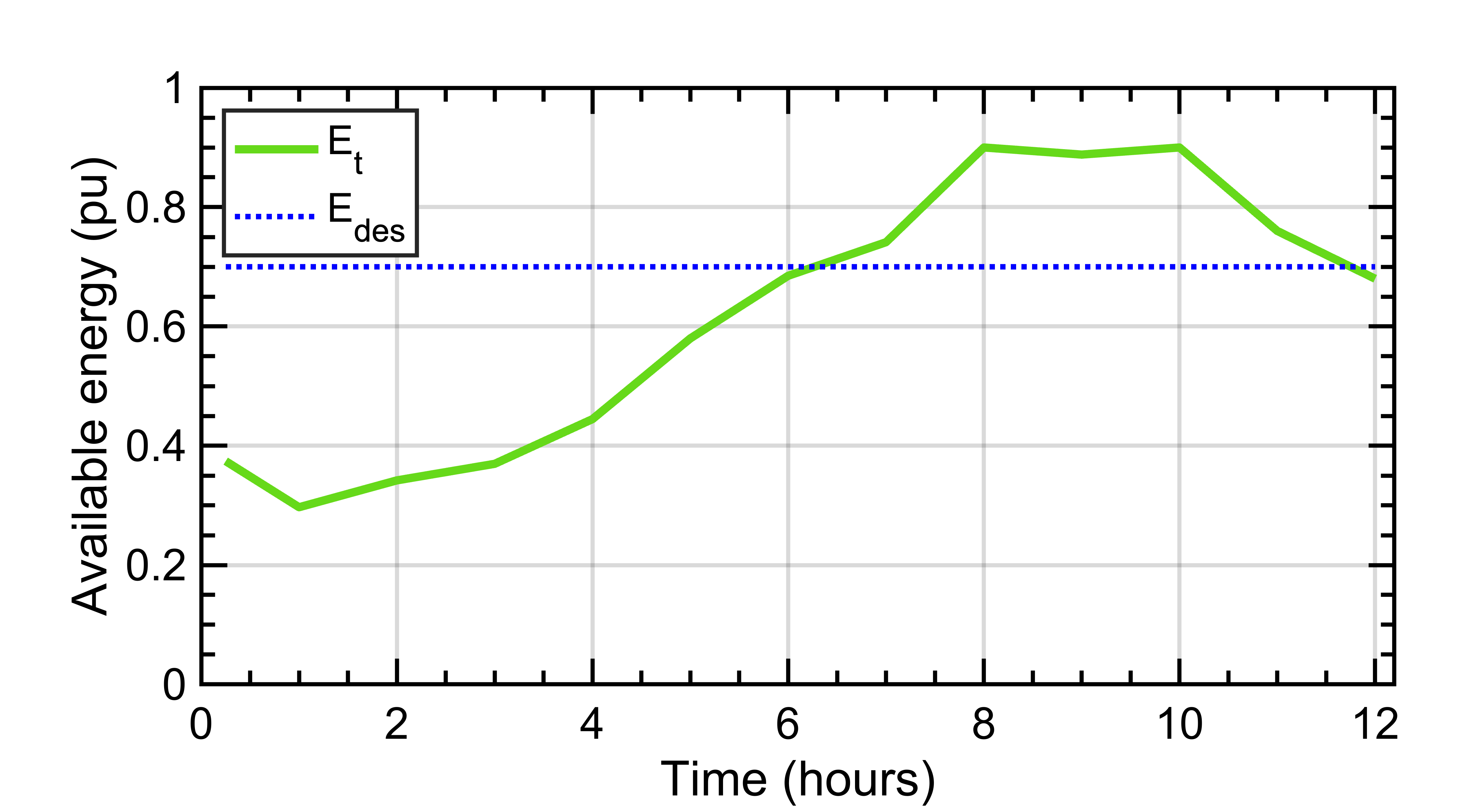} 
            \caption{}
            \label{regret}
	\end{subfigure}
	\caption{ \small (a) Charging power ($P_{\text{bat,t}}$) profile with V2G tariff ($\alpha_t$). Positive $P_{\text{bat,t}}$ implies charging and negative value refers to discharging. (b) Available energy ($E_t$) in the battery during V2G charging session.}
\end{figure}
\subsubsection{Smart Charging Profiles}
Figures (\ref{sensi}) and (\ref{regret}) show the charging power profile and the available state of energy of the EV battery, respectively. Note that the charging/discharging profile depends on the V2G tariff price, selecting to discharge, i.e., sell energy to the grid retailer or parking lot manager when the electricity price is high and charge when the electricity price is low.  

\section{Conclusion}
In this paper, we proposed a data-driven, user-centric V2G methodology that balances battery degradation and V2G revenue using multi-objective optimization and ICNN-based battery degradation modeling. The partial convexity of our data-driven model ensures computational efficiency and global optimality, while numerical simulations demonstrate its high accuracy ($R^2$= 0.974) and adaptability to unseen data. Future work will focus on the experimental validation of our approach in real V2G scenarios. In such a setting, an interesting extension would be to use data-driven methods to predict the preferences of the prosumers. Such predictions can be leveraged by the grid retailer or parking lot managers to improve their energy management strategies further.

\bibliographystyle{IEEEtran}
\bibliography{references}

\end{document}